\documentclass[reprint,superscriptaddress,
amsmath, amssymb,aps, prl,notitlepage,longbibliography,floatfix,
nofootinbib 
]{revtex4-2}

\usepackage{graphicx}   
\usepackage[
colorlinks=true,        
citecolor=blue,         
linkcolor=blue,         
urlcolor=blue           
]{hyperref}             
\usepackage{bm}         
\usepackage{xcolor}     
\newcommand{\nc}{\newcommand*}

\nc{\al}{\alpha}
\nc{\s}{\sigma}
\nc{\kp}{\kappa}
\nc{\dt}{\delta}
\nc{\Dt}{\Delta}
\nc{\Ld}{\Lambda}
\nc{\p}{\partial}
\nc{\Gm}{\Gamma}
\nc{\om}{\omega}
\nc{\Om}{\Omega}
\nc{\rd}{\mathrm{d}}
\nc{\Eq}[1]{Eq.~\eqref{#1}}     
\nc{\Fig}[1]{Fig.~\ref{#1}}     
\nc{\Table}[1]{Table~\ref{#1}}  
\nc{\Sec}[1]{Sec.~\ref{#1}}     
\nc{\Msun}{M_\odot}             
\nc{\Ogw}{\Omega_{\mathrm{GW}}}
\nc{\gpcyr}{\mathrm{Gpc}^{-3}\,\mathrm{yr}^{-1}}
\nc{\lvc}{LIGO/Virgo} 
\nc{\SNR}{\mathrm{SNR}} 
\nc{\rhoGW}{\rho_{\mathrm{GW}}}
\nc{\vd}{\vec{d}}
\nc{\av}[1]{\langle #1 \rangle} 
\nc{\km}{\mathrm{km}}
\nc{\Mpc}{\mathrm{Mpc}}
\nc{\Tobs}{T_{\mathrm{obs}}}
\nc{\fyr}{f_{\mathrm{yr}}}
\nc{\hii}{\rm H~{\textsc {ii}}}
\nc{\addref}{[\textcolor{red}{add ref}] } 
\nc{\eg}{\textit{e.g.~}}
\nc{\app}{\approx}
\nc{\hf}{\frac{1}{2}}
\nc{\discuss}{\textcolor{red}{Add discussion here!}}
\nc{\red}[1]{\textcolor{red}{#1}}
\nc{\hp}{h_+} 
\nc{\hc}{h_{\times}} 
\nc{\mbh}{M_{\rm BH}}
\nc{\mdisk}{M_{\mathrm{disk}}}
\nc{\rdisk}{r_{\mathrm{disk}}}
\nc{\mc}{M_{\mathrm{c}}}

\begin{document}
	
\title{How Fast do Black Holes Spin?}



\author{Chen Yuan}
\email{chenyuan@tecnico.ulisboa.pt}
\affiliation{CENTRA, Departamento de Física, Instituto Superior Técnico – IST, Universidade de Lisboa – UL, Avenida Rovisco Pais 1, 1049–001 Lisboa, Portugal}
\author{Vitor Cardoso} 
\affiliation{Center of Gravity, Niels Bohr Institute, Blegdamsvej 17, 2100 Copenhagen, Denmark}

\date{\today}

\begin{abstract}
We present a systematic study of the spin evolution of accreting black holes. Using the \texttt{BHAC} code for general relativistic magnetohydrodynamic simulations, we conduct a total of 500 high resolution simulations of magnetized accretion disks around Kerr BHs with spin parameters ranging from $\chi=0$ to $\chi = 0.99$. 
The baseline configuration follows the default Fishbone–Moncrief torus setup in \texttt{BHAC}. 
We further explore the effects of: I) introducing multiple poloidal magnetic loops, II) the initial position of the torus, III) magnetic field strength and IV) radial loop modulation. 
From each simulation we extract the mass flux and angular momentum flux through the horizon to compute the time-averaged spin evolution rate after the system becomes quasi-stationary.
The equilibrium spin ranges from $\chi_{\rm eq}\simeq0.88$ in the fiducial model to $\chi_{\rm eq}\simeq0.98$ for weak-flux model. Applied to Sgr~A$^*$, even an optimistic estimate for rapid spin evolution gives a timescale longer than $10^{13}\,$yr, indicating an earlier growth or merger history.
These trends are consistent with the picture that larger net magnetic flux threading the horizon enhances jet-torques and reduces the equilibrium spins, whereas weak net flux yields near-maximal spins. Our results provide a pathway to interpret the high spins reported for X-ray binaries and active galactic nuclei and they sharpen the astrophysical viability of near-horizon high-energy processes that rely on near-extremal black holes.
\end{abstract}

\maketitle

\noindent {\bf \em Introduction.} 
Accreting black holes (BHs) are remarkably efficient engines in nature, converting rotational and gravitational binding energy into radiation. 
Across both stellar-mass and supermassive BHs, X-ray reflection spectroscopy and thermal continuum fitting have yielded numerous spin measurements, with many sources showing near-maximal values. As summarized by recent reviews, these two methods provide the most mature and widely applied spin diagnostics and collectively reveal a substantial population of rapidly rotating BHs \cite{Garcia:2013lxa,McClintock:2013vwa,Reynolds:2020jwt,Draghis:2022ngm} (the dimensionless spin of BHs in vacuum General Relativity satisfies $\chi <1$ in the units adopted in this work). For example, the well-studied X-ray binary Cygnus X-1 has been shown via the continuum-fitting method to have a dimensionless spin 
%
%
$\chi > 0.983$ ($3\sigma$) \cite{Gou:2013dna}. Independently, X-ray reflection spectroscopy yields a comparable value of $\chi \simeq 0.97$ \cite{Fabian:2012kv}. These consistently high spin measurements, persisting across methods, lead to a critical constraint on theoretical models of BH spin evolution.
Additionally, the high-mass X-ray binary LMC X-1 has been estimated to have $\chi\simeq 0.92$~\cite{Gou:2009ks}. Meanwhile, GRS 1915$+$105, when analyzed with updated system parameters, still permits near-maximal spin values (best fit $\chi \simeq 0.86$ in one dataset, and up to the model limit $\chi \simeq 0.99$), although the uncertainty envelope widens substantially~\cite{Mills:2021dxs}. Finally, a number of highly spinning BHs have been reported in the X-ray band~\cite{Draghis:2022ngm} (see also their Fig. 19). Together, these results illustrate both the accumulating evidence for highly spinning BHs and the sensitivity of spin estimates to system parameters and disk modeling assumptions \cite{Parker:2015fja,Miller-Jones:2021plh,Zhao:2021bhj,Draghis:2022ngm}.

High spins are also inferred for several nearby active galactic nuclei via relativistic reflection spectroscopy.  For example, the Seyfert galaxy {MCG–6-30-15} yields a formal constraint $\chi > 0.987$ (90\% confidence) under the assumption of no emission from within the innermost stable circular orbit \cite{Brenneman:2006hw}.  Similarly, NuSTAR and XMM-Newton observations of NGC 1365 demonstrate that even when accounting for strong line-of-sight absorption, the spin remains high, with $\chi > 0.97$ in all independent epochs \cite{Risaliti:2013cga}.  More recently, high-density disk reflection modeling has shown that by relaxing the conventional assumption of fixed disk density the required extreme iron abundances can be significantly reduced, while the conclusions of rapid rotation are largely preserved \cite{Jiang:2019ztr}.  Together, these observational findings reinforce the case for a substantial population of rapidly spinning supermassive BHs.

Within the classical Novikov--Thorne (NT) thin disk framework, the spin of a Kerr BH regulates the radiative efficiency through the innermost stable circular orbit (ISCO) \cite{Page:1974he,Thorne:1974ve}. The NT model further assumes that only photons carry stress-energy away from the disk, explicitly excluding extended magnetic fields. Within this framework, the BH spin evolves through a competition between the accretion torque at the horizon and the capture of returning radiation, yielding the well known Thorne limit of $\chi_{\mathrm{eq}} \simeq 0.998$ \cite{Thorne:1974ve}.

Beyond the classical NT picture, earlier analytic arguments and subsequent general relativistic magnetohydrodynamics (GRMHD) simulations indicate that magnetic stresses persist across and even inside the ISCO, altering the angular-momentum flux through the horizon \cite{Agol:1999dn,Noble:2010mm,Noble:2011wa}.
These departures from the NT assumptions lead to a modified $\chi_{\rm eq}$ that depends on disk thickness and magnetic geometry \cite{Penna:2010hu,Tchekhovskoy:2011zx}.
For the magnetically arrested disk (MAD) case, large scale vertical magnetic flux accumulates and rotational energy can be extracted through the Blandford–Znajek process \cite{Blandford:1977}, leading to relativistic jets whose torques tend to spin the BH down \cite{Tchekhovskoy:2011zx,Avara:2015kna}.
In contrast, in weak flux, ``standard and normal evolution'' (SANE) disks, jet-torques are small and the resulting equilibrium spin is correspondingly higher than in MAD states.
On the other hand, in advection dominated/radiatively inefficient accretion flow (ADAF/RIAF), analytic and GRMHD studies of advection-dominated/radiatively inefficient flows show that, owing to the suppression of radiative capture and the persistence of magnetic stress inside the ISCO, the resulting equilibrium spin deviates from the NT limit~\cite{Popham:1998iw,Narayan:2012yp}.

In GRMHD flows without radiative losses (SANE-like/thick disks), simulations of thick, radiatively inefficient disks yield $\chi_{\rm eq}\simeq 0.9$ \cite{Gammie:2003qi}, significantly below the Thorne limit. Semi-analytic ADAF models find $\chi_{\rm eq}\simeq 0.96$ in the absence of jets and $\chi_{\rm eq}\simeq 0.93$ once jet braking is included \cite{Benson:2009kx}. 
Super-Eddington ``slim'' disks can push the equilibrium spin even closer to unity (e.g., $\chi_{\rm eq}=0.9994$ for ten times the Eddington limit and viscosity parameter $\eta=0.01$) \cite{Sadowski:2011ka}. 
By contrast, GRMHD simulations of MADs threaded by strong vertical magnetic flux show that BZ jets can spin-down the BH to very low equilibrium spins, $\chi_{\rm eq}\sim 0.04-0.07$ \cite{Tchekhovskoy:2011zx,Lowell:2023kyu}. Semi-analytic population models link different fueling modes to characteristic spin distributions: prolonged coherent accretion drives spins toward near-maximal values, whereas chaotic accretion episodes and jet-torques produce broader, lower-spin populations \cite{Sesana:2014bea}.

Highly spinning BHs provide a unique way to probe near-horizon, strong-field dynamics, hence to test General Relativity. Close to extremality, the Penrose mechanism allows particle collisions arbitrarily close to the horizon to achieve formally unbounded center-of-mass energy for appropriately
fine-tuned geodesics~\cite{Banados:2009pr,Berti:2009bk,Schnittman:2014zsa,Mummery:2025cog,Brito:2015oca}. Such high energies -- larger than what can be achieved with Earth-bound accelerators -- might possibly provide hints for new physics. More recently, exciting developments in semiclassical gravity, also place extremal BHs on the spotlight: on one hand, it has been argued that the length of the throats of BHs sufficiently close to
extremality exhibits large quantum variance~\cite{Iliesiu:2020qvm}. On the other hand, stringy or quantum corrections to general relativity are expected to generate higher-curvature terms in the gravitational action, which cause extremal rotating BHs to have curvature singularities on their horizon~\cite{Horowitz:2023xyl}. In other words, extremal BHs are intrinsically quantum objects and sensitive probes of new physics. 

These considerations warrant a systematic GRMHD study of BH spin evolution across accretion states with different magnetic-flux supply, with the goal of quantifying how electromagnetic stresses
and horizon-scale flux regulate the approach to (or departure from) near-extremal equilibrium. Motivated by the above context, we present a systematic GRMHD study of steadily accreting BHs, with the goal of quantifying the time-averaged spin evolution rate $\langle \dot{\chi} \rangle$ as a function of the BH spin across a controlled survey of magnetic topologies, magnetization, torus position and fluid properties.

It is important to state the scope of this Letter explicitly. 
The goal is not to determine a universal equilibrium spin for every possible astrophysical accretion flow. 
The spin evolution of a real BH depends on its disk thermodynamics, magnetic flux supply, radiative efficiency, and possible transitions between SANE and MAD states. 
Instead, motivated by the high spins inferred in several X-ray binaries and AGN, and by the theoretical importance of near extremal BHs, we ask a more focused question: under what magnetized accretion conditions can accretion produce or maintain high BH spins?

\noindent {\bf \em Method.} 
\label{sec:method}
Our simulations are carried out with the open‐source GRMHD code \texttt{BHAC}~\cite{Porth:2016rfi,Olivares:2019dsc,EventHorizonTelescope:2019pcy}, under the axisymmetric approximation in a fixed Kerr spacetime.
In the geometric units we adopt here, with the speed of light and Newton's constant $c=G=1$. The Kerr BH spacetime is characterized by its mass $M$ and the dimensionless spin $\chi$. We solve the GRMHD equations using modified Kerr–Schild (MKS) coordinates with logarithmic radial coordinate to concentrate resolution near the event horizon \cite{Gammie:2003rj}. 
The computational domain extends from the horizon radius out to $2500\,M$, where $M$ is the mass of the BH and $\theta\in[0,\pi]$ spans the full polar angle.
In our $(\ln r,\theta)$ grid we employ a base resolution of $128\times128$, and enable three levels of adaptive mesh refinement. This yields a finest‐level effective resolution of $512\times512$ in $(\ln r,\theta)$. Models Fiducial, I, and III are evolved to $5000M$, Model~II to $10000M$ because its torus starts farther out, and the high-spin Model~IV runs used to locate its turnover to at least $2700M$. The standard duration corresponds to
\begin{equation}
5000M = 5000\frac{GM}{c^3}
\simeq 28.5
\left(\frac{M}{10^8M_\odot}\right)\ {\rm days}.
\end{equation}
Each simulation is initialized with a Fishbone–Moncrief (FM) equilibrium torus~\cite{FM1976}. For the configuration inside the torus, the torus is filled with an ideal fluid of adiabatic index $\Gamma=4/3$.
The fluid is bound between an inner edge $r_{\rm in}$ and an outer edge $r_{\rm out}$, with maximum density at $r_{\max}$.
Following the standard FM construction, we first fix $(r_{\rm in}, r_{\max})$. Then we enforce a constant specific angular momentum throughout the torus, $\ell = {u_{\phi}}/{u_{t}}$, where $u_{\phi}$ and $u_{t}$ are the azimuthal and time components of the fluid four‑velocity in Boyer–Lindquist coordinates. The angular momentum parameter $\ell$ is determined so that the density is peaked at $r_{\max}$.
The equilibrium is described by the ``effective potential''
\begin{equation}
    W(r,\theta) \equiv -\ln\!\bigl[-u_t(r,\theta)\bigr] - C\,,
\end{equation}
where $C$ is fixed by setting $W(r_{\rm in},\pi/2)=0$ so that $r_{\rm in}$ denotes the inner bound.
In the implementation we compute the enthalpy excess 
$\ln h(r,\theta) \equiv W(r,\theta)-W(r_{\rm in},\pi/2)$
and identify torus material wherever $\ln h>0$, while outside this region the atmosphere floors are applied to avoid vacuum.
On the other hand, the outer edge $r_{\rm out}$ is identified as the equatorial radius $r>r_{\max}$ where $\ln h=0$ is first encountered, so that the enthalpy outside this region returns to unity and the fluid becomes unbound. Together, $(r_{\rm in},\,r_{\max},\,r_{\rm out})$ define the radial extent of a bound equilibrium FM torus.

\begin{table}[ht!] 
\scriptsize
\setlength{\tabcolsep}{1.5pt}
\caption{Model configurations used in this work. Except for the parameters listed here, all other numerical settings are identical to the Fiducial model. For Model~III the last column gives a transition interval.}
\label{tab:models}
\begin{tabular}{lccccc}
\hline\hline
Models & $(r_{\rm in},\,r_{\max})$ & $\beta_{\max}$ & $N_{\rm loop}$ & $\lambda_r$ & $\chi_{\rm eq}$ 
\\
\hline
Fiducial                          & $(6,12)M$   & $100$ & $1$ & $0$ & 0.88 \\
I (multiple poloidal loop)  & $(6,12)M$  & $100$ & $3$ & $0$ & 0.87 \\
II (shifted torus position) & $(10,20)M$ & $100$ & $1$ & $0$ & 0.78 \\
III (high magnetization)    & $(6,12)M$  & $10$  & $1$ & $0$ & \shortstack{$0.37$--$0.71$} \\
IV (radial loop modulation) & $(6,12)M$  & $100$ & $1$ & $4M$ & 0.98 \\
\hline
\end{tabular}
\end{table}
The pressure of the magnetic field is $p_{\rm mag}=b^2/2$ with $b^\mu$ the comoving magnetic field 4-vector. 
$\beta$ is therefore defined as $\beta = p/p_{\text{mag}}$. Then we choose a free parameter $\beta_{\max}$, defined as the torus pressure
maximum ($p_{\max}$) over the magnetic pressure maximum ($b_{\max}^2/2$). And a rescaling factor
\begin{equation}
{\cal N} \;=\; \sqrt{\frac{2\,p_{\max}}{\,b^2_{\max}\,\beta_{\max}}}\, ,
\label{eq:norm}
\end{equation}
is used to rescale the magnetic field such that $b^\mu \to {\cal N}\,b^\mu$. One can check that, in such a notation, the maximum is still $\beta_{\max}\to 2\,p_{\max}/({\cal N}^2\,b^2_{\max})
\;=\;\beta_{\max}$.
In our fiducial model, we set $\beta_{\max}=100$ corresponding to a weak magnetic field configuration.
On the other hand, the initial magnetic field is generated from an azimuthal vector potential $A_\phi$. 
Let $[\cdot]_+\equiv\max(\cdot,0)$ denote the positive-part operator.
Inside the FM torus one sets
\begin{equation}
A_\phi(r,\theta) \;=\; \bigl[\rho(r,\theta)-\rho_{\rm cut}\bigr]_+ \;
f_\theta(\theta) f_r(r)\, ,
\label{eq:Aphi-core}
\end{equation}
with $\rho_{\rm cut}$ a small threshold ($0.2$ in code units) localizing $A_\phi$ to dense torus material.
The factors $f_\theta$ and $f_r$ encode the field topology:
\begin{align}
f_\theta(\theta) &=
\begin{cases}
1, & N_{\rm loop}=1,\\[3pt]
\cos\!\bigl[(N_{\rm loop}-1)\,\theta\bigr], & N_{\rm loop}>1,
\end{cases}
\label{eq:ftheta}
\\
f_r(r) &= 
\begin{cases}
1, & \lambda_r=0,\\[3pt]
\sin\!\Bigl( \dfrac{2\pi\,(r-r_{\rm in})}{\lambda_r}\Bigr), & \lambda_r>0,
\end{cases}
\end{align}
where $N_{\rm loop}$ sets the number of poloidal magnetic flux loops in the torus and $\lambda_r$ introduces a radial modulation of the loops.

Finally, to seed the MRI, we apply a $4\%$ random perturbation to the gas pressure within the torus. Away from the torus, a low‑density ``atmosphere’’ is imposed with floor values to avoid vacuum states \cite{Porth:2016rfi}
\begin{equation}
  \rho_{\rm floor}(r) \;=\; 10^{-5}\,r^{-3/2},
  \qquad
  P_{\rm floor}(r) \;=\; \frac{10^{-7}}{3}\,r^{-5/2}.
\end{equation}

After the quasi‐stationary phase, we extract the mass and angular‐momentum fluxes through the horizon,
\begin{equation}
\begin{aligned}
\dot M_h &= -\int T^{r}{}_{t} \,\sqrt{-g}\,d\theta\,d\varphi, \\
\dot J_h &= -\int T^{r}{}_{\varphi}\,\sqrt{-g}\,d\theta\,d\varphi,
\end{aligned}
\end{equation}
with the stress–energy tensor of a magnetized fluid defined as
\begin{equation}
    T_{\mu\nu} = (\rho + P + \rho\epsilon + b^2)\,u_\mu u_\nu + \Bigl(P + \tfrac12 b^2\Bigr)\,g_{\mu\nu} - b_\mu b_\nu.
\end{equation}
We then compute the instantaneous time evolution of the dimensionless spin
\begin{equation}
\dot\chi = \frac{\dot J_h}{M^2} \;-\; 2\,\chi\,\frac{\dot M_h}{M}\,.
\end{equation}

\begin{figure*}[t]
\centering
  \includegraphics[width=0.9\linewidth]{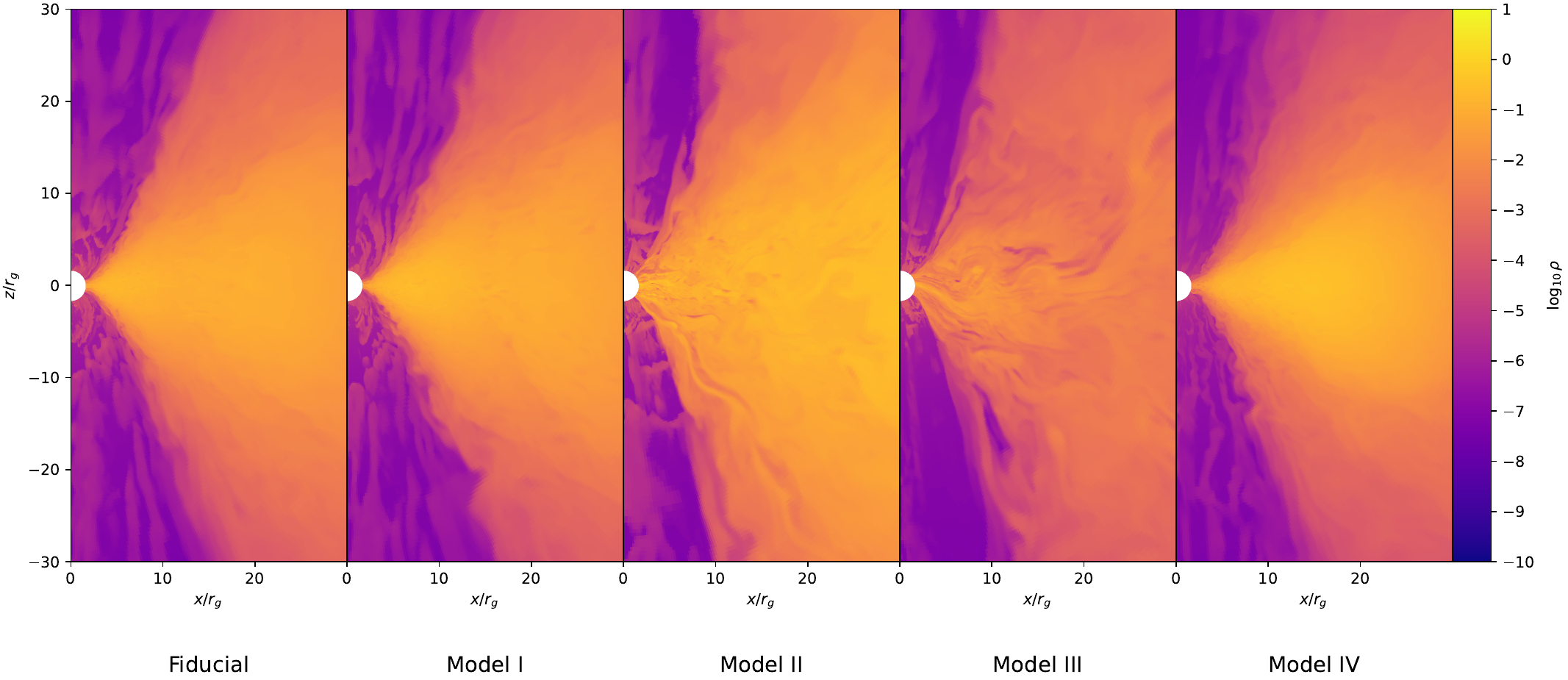}
    \caption{The rest-mass density for all five models in the quasi-stationary phase for $\chi=0.7$. The characteristic toroidal structure of the accretion flow and the low-density funnel regions comprising the jet are shown. }
  \label{fig:10panels}
\end{figure*}
For each configuration in Table~\ref{tab:models}, we simulate a set of 100 spins uniformly spaced over $\chi\in[0,0.99]$. All other parameters and numerical settings are identical to the Fiducial model, allowing the effect of each modified parameter to be assessed in a controlled manner. 
After the system enters a quasi‐stationary state, we compute the horizon–integrated fluxes of mass and angular momentum and evaluate the instantaneous $\dot\chi$. We report its time median and 5th--95th percentiles.

%
\noindent {\bf \em Results.} \label{sec:results}
Fig.~\ref{fig:10panels} shows snapshots of the rest-mass density of all five models during this phase. It can be seen that, in all the models, the characteristic toroidal structure of the accretion flow is preserved, with the density maximum remaining near $r_{\rm max}$. A low–density funnel region comprising the jet is evident.

\begin{figure*}
  \includegraphics[width=0.45\linewidth]{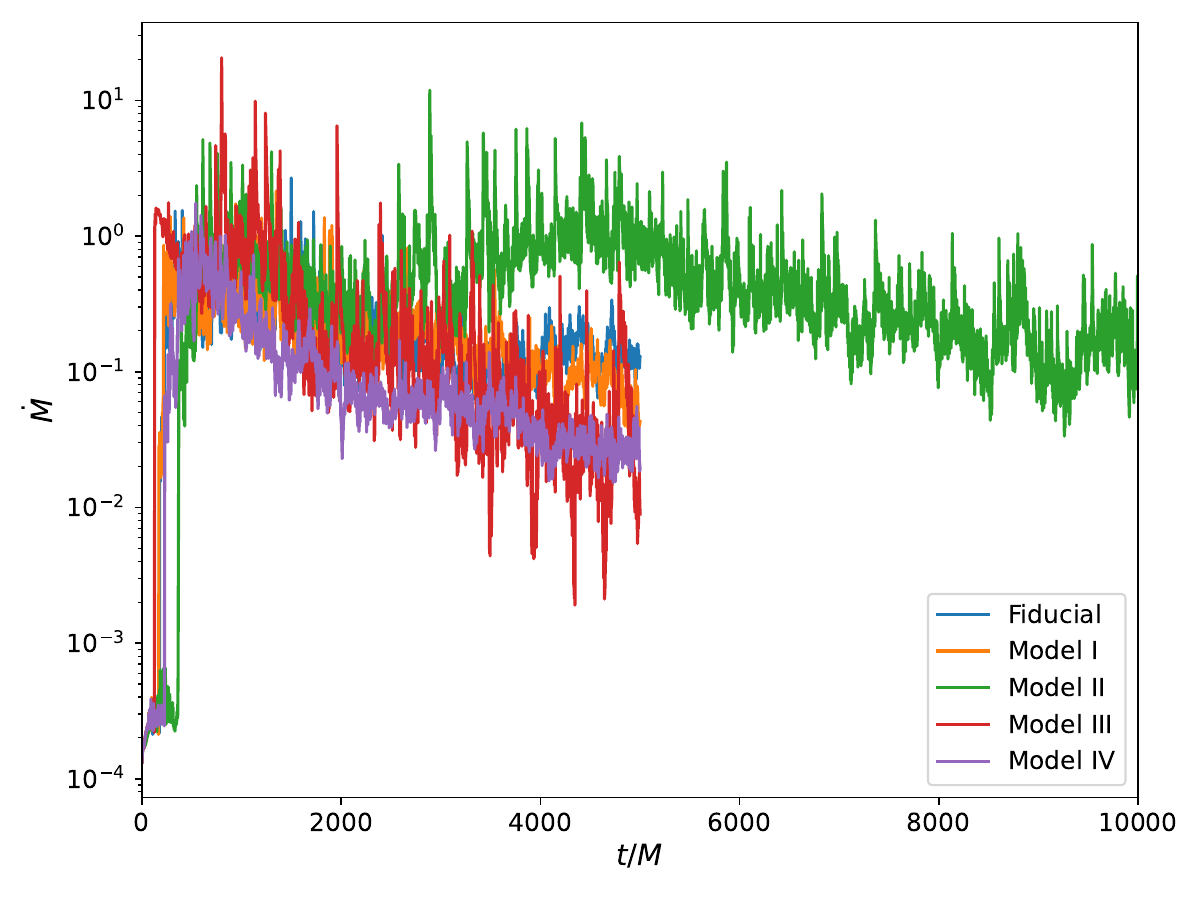}
    \includegraphics[width=0.45\linewidth]{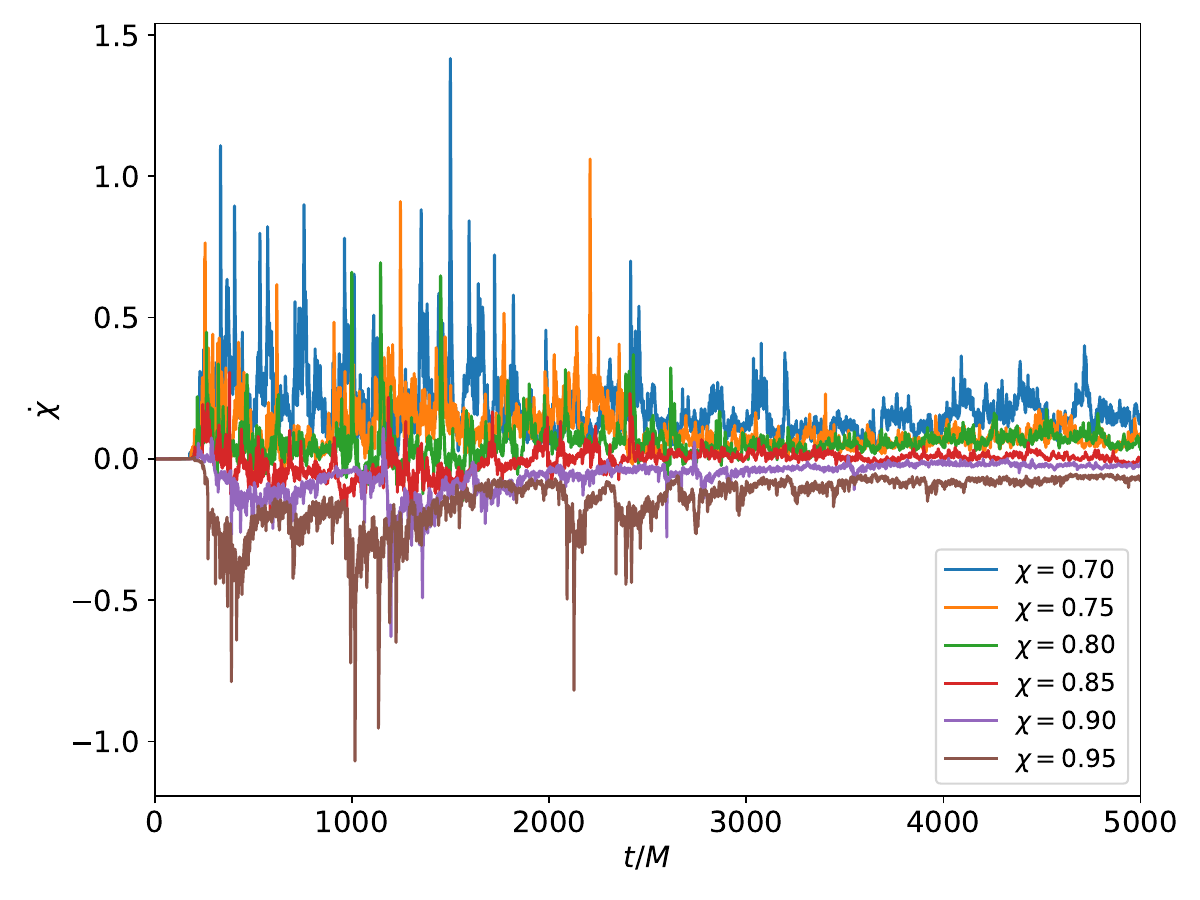}
  \caption{\textbf{Left panel}: The time evolution of $\dot{M}$ for different models, fixing $\chi=0.7$. \textbf{Right panel:} The time evolution of $\dot{\chi}$ for different BH spins, given by the fiducial model.}
  \label{fig:m-chi-t}
\end{figure*}
In Fig.~\ref{fig:m-chi-t} we show the time series of the mass accretion rate, $\dot{M}$, and the instantaneous spin evolution rate, $\dot{\chi}$ over $0 \le t \le 5000\,M$.
After $t\sim100\,M$, the fluid flux crosses the horizon and the accretion process begins.
Once the flow reaches the quasi-stationary phase, the torus is progressively depleted and, in a viscosity-dominated regime with approximately constant effective transport, the mass accretion rate exhibits an approximately exponential decay in time. This is also reflected in the $\dot{\chi}$ panel: for the fiducial model, the large early-time fluctuations are reduced once the flow enters the quasi-stationary phase at $t\gtrsim3000M$.
For near‑extremal spins, it can be seen that $\dot{\chi}$ fluctuates almost entirely below zero, indicating that the BH is in a spin-down regime.
We use $[4000M,5000M]$ for the Fiducial and Model~I curves, $[7000M,8000M]$ for Model~II, and $[2000M,3000M]$ for Model~III. Repeating the calculation with shifted, fixed-length windows leaves the inferred turnovers unchanged within the sampled spin spacing of $0.01$; all windows place the Model~IV equilibrium near $\chi_{\rm eq}\simeq0.98$.
For Model~III, which is initialized with a much stronger magnetic field ($\beta_{\max}=10$), the evolution proceeds significantly faster. The enhanced magnetic stresses lead to more efficient angular momentum transport and a rapid draining of the initial torus. As a consequence, beyond $t \gtrsim 3000M-4000M$ the disk mass declines steeply. In this late phase the horizon flux is no longer representative of a massive, geometrically thick accretion flow. To avoid this regime, for
Model~III we identify an earlier quasi–steady window $t \in [2000M,\,3000M]$, during which the accretion rate and magnetic flux have already saturated while the bulk of the torus is still present and clearly visible.

\begin{figure}
  \centering
  \includegraphics[width=0.9\linewidth]{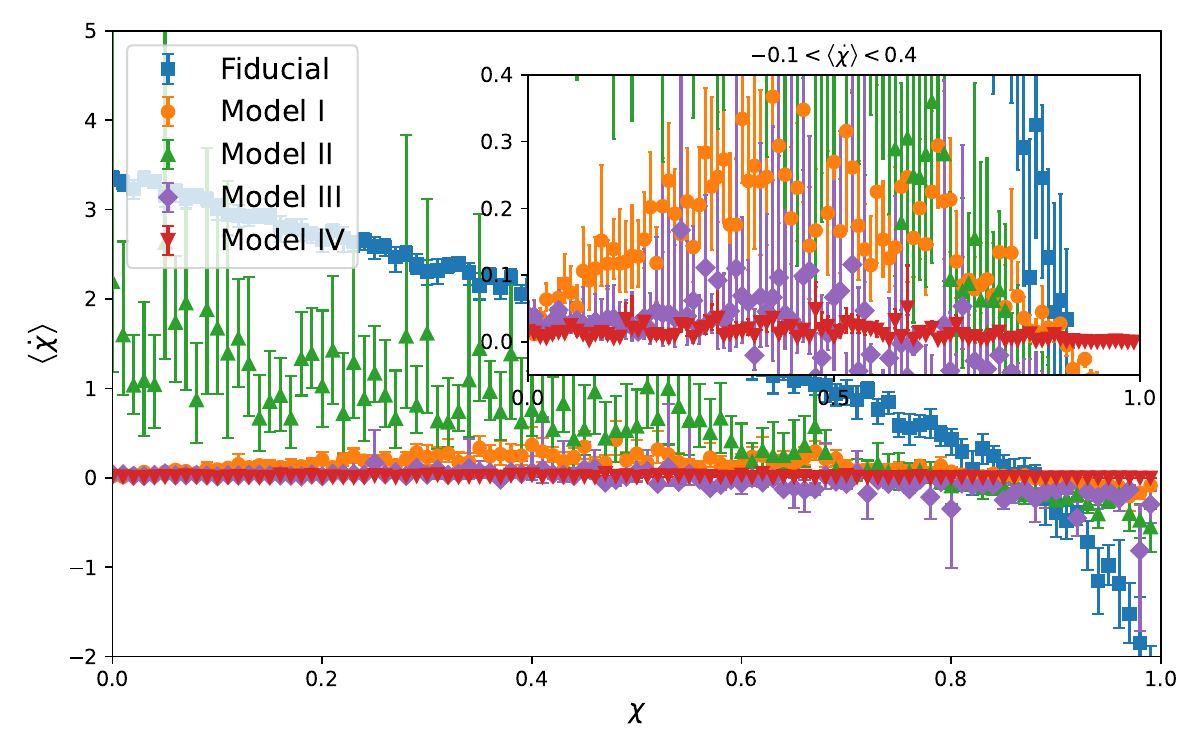}
  \caption{The median value and the $90\%$ C.L. region of $\dot\chi$ versus BH spin $\chi$ for five models. These results are given in the code units of \texttt{BHAC}. Physical units follow from Eq.~(\ref{rescale}).}
  \label{fig:chidot}
\end{figure}

Our main results are summarized in Fig.~\ref{fig:chidot}, where we show the time‑averaged spin evolution rate, $\left\langle \dot{\chi} \right \rangle$, as a function of the BH spin $\chi$ for all the models.
A key feature of interest is the turning point (zero-crossing) of $\langle \dot{\chi} \rangle$, which indicates the transition of the BH spin evolution from spin-up to spin-down. 
Our setup for the fiducial model is chosen to closely match the 2D FM torus configuration of \cite{Gammie:2003qi} ($r_{\rm in}=6M$, $r_{\max}=12M$, $\Gamma=4/3$, weakly magnetized poloidal field). The equilibrium spin of our fiducial model is $\chi_{\rm eq}\simeq 0.88$, lower than the value $\chi_{\rm eq}\simeq 0.93$ reported by \cite{Gammie:2003qi}. Such a difference may be attributed to the different definition of $\beta$. \cite{Gammie:2003qi} uses a minimum gas-to-magnetic pressure ratio of $100$, while our parameter $\beta_{\max}=100$ is defined through $p_{\max}/(b_{\max}^2/2)$.
It is useful to contrast our results with recent studies of the strongly magnetized MAD regime. 
Long-lived MAD simulations show that large-scale magnetic flux can drive efficient Blandford--Znajek outflows and spin the BH down \cite{Chatterjee:2022mxg,Narayan:2021qfw,Lowell:2025ris}. 
For example, luminous thin MADs can reach a low equilibrium spin, $a_{\rm eq}\sim0.3$, while hot MAD flows can remove more angular momentum through jets than is supplied by accretion. 
Flux eruptions in MADs further lead to intermittent and non-axisymmetric angular-momentum transport. 
These results are complementary to our analysis. 
They describe the strong flux limit in which high spins are efficiently braked. However, our main goal is to determine under what conditions high or near-extremal spins remain viable.  Our strongly magnetized Model III shows the same tendency: increasing the net magnetic flux lowers the equilibrium spin.

The turnover depends strongly on magnetic configuration: Model~I gives $\chi_{\rm eq}\simeq0.87$, Model~III has a $0.37$--$0.71$ transition interval rather than a unique equilibrium, and weak-flux radial modulation in Model~IV permits $\chi_{\rm eq}\simeq0.98$. This trend is consistent with stronger coherent horizon flux producing larger jet torques~\cite{Blandford:1977,Tchekhovskoy:2011zx}.

We have also performed additional robustness tests of the high-spin branch. 
Varying $\lambda_r$, radial phase $\phi_r$, $\beta_{\max}$, $r_{\rm in}$, $r_{\rm max}$, and the cutoff density in the potential, 
\begin{equation}
 A_\phi \propto
 \max(\rho-\rho_{\rm cut},0)
 \sin\left(
 2\pi\frac{r-r_{\rm in}}{\lambda_r}+\phi_r
 \right),
\end{equation}
we explored $\lambda_r=2.5M,3M,3.5M,4M$, 
$\phi_r=\pi/4,\pi/2,3\pi/4,\pi$, 
$\beta_{\max}=75,100,125,150,175$, 
$r_{\rm max}=14M,16M$ with $r_{\rm in}=6M$, 
$r_{\rm in}=8M,10M$ with $r_{\rm max}=12M$, 
and $\rho_{\rm cut}=0.3,0.4,0.5$. 
These variations generally preserve a high-spin turnover above $\chi\simeq0.9$, showing that the high-spin branch is not tied to one fine-tuned initial condition.

\noindent {\bf \em Astrophysical implications.} \label{sec:implications}
To discuss the astrophysical implications of the results, one has to convert the data output by \texttt{BHAC} from code units into physical units \cite{Younsi2012,Younsi2023}. The rescaling of the data is done by choosing a physical accretion rate, (see e.g. \cite{Yuan:2025fde})
\begin{equation}
\begin{aligned}
\dot M^{\rm phys} &= f_{\rm Edd}\,\dot M_{\rm Edd},\\
\dot M_{\rm Edd} &\approx 2\times10^{-8}
\left(\frac{M}{M_{\odot}}\right)M_{\odot}\,\mathrm{yr}^{-1}.
\end{aligned}
\end{equation}
where $\dot{M}_{\rm Edd}$ is the Eddington accretion rate and $f_{\rm Edd}$ represents the dimensionless accretion rate. Then the stress-energy tensor is rescaled as
\begin{equation}
T_{\mu\nu}^{(\rm phys)}=
\frac{\mathcal{M}\,c^{3}}{r_{g}^{3}}\;
T_{\mu\nu}\,,
\end{equation}
with $r_{g} \equiv {G M}/{c^{2}}$ and the rescale factor is given by $\mathcal{M}\equiv \dot{M}^{\rm phys} (r_g/c)$. 
As a result, the code unit $\dot{\chi}$ rescales as 
\begin{equation}\label{rescale}
    \dot{\chi}^{\rm phys} = \frac{\mathcal{M}}{M(r_g/c)}\dot{\chi} =  2\times10^{-8} f_{\rm Edd} \dot{\chi}~\mathrm{yr}^{-1}.
\end{equation}

An immediate application is Sgr~A$^*$. EHT modeling gives $\dot M\simeq(5.2\text{--}9.5)\times10^{-9}M_\odot\,{\rm yr}^{-1}$~\cite{EventHorizonTelescope:2022xqj}. Even taking an optimistic case for rapid spin evolution, with only $\Delta\chi=0.1$ and a large value $|\dot\chi|=4$ from our survey, Eq.~(\ref{rescale}) gives a required time longer than $10^{13}\,$yr. Thus, if the reported estimate $\chi\simeq0.90\pm0.06$ is confirmed~\cite{Daly:2023sgr}, the present accretion flow cannot yield that spin, indicating an earlier high accretion or merger epoch~\cite{Wang:2024sgrmerger}.

We were driven primarily by foundational questions regarding BHs {\it extremely} close to extremality: can they work as particle accelerators, or behave as intrinsically quantum objects~\cite{Banados:2009pr,Iliesiu:2020qvm}?
The accretion spin limit is therefore a magnetic-state-dependent attractor, not a universal constant. Combining spin with the accretion and magnetic state tests whether the current flow built it or whether it is a fossil of earlier growth or mergers.

\noindent {\bf \em Acknowledgments.} 
%
We are grateful to Juri Poutanen for correspondence on literature on X-ray observations. We also thank Ziri Younsi and Oliver Porth for useful comments on \texttt{BHAC}.
The Center of Gravity is a Center of Excellence funded by the Danish National Research Foundation under grant No. DNRF184.
We acknowledge support by VILLUM Foundation (grant no. VIL37766).
V.C.\ is a Villum Investigator.  
We acknowledge financial support provided under the European Union’s H2020 ERC Advanced Grant “Black holes: gravitational engines of discovery” grant agreement no. Gravitas–101052587. Views and opinions expressed are however those of the author only and do not necessarily reflect those of the European Union or the European Research Council. Neither the European Union nor the granting authority can be held responsible for them. 
We thank the Fundação para a Ciência e Tecnologia (FCT), Portugal, for the financial support to the Center for Astrophysics and Gravitation (CENTRA/IST/ULisboa) through grant No. UID/PRR/00099/2025 (https://doi.org/10.54499/UID/PRR/00099/2025) and grant No. UID/00099/2025 (https://doi.org/10.54499/UID/00099/2025)
This project has received funding from the European Union's Horizon 2020 research and innovation programme under the Marie Sklodowska-Curie grant agreement No 101007855 and No 101131233.
This work is supported by Simons Foundation International \cite{sfi} and the Simons Foundation \cite{sf} through Simons Foundation grant SFI-MPS-BH-00012593-11.
\bibliography{./ref}

\end{document}